# Electronic structure of buried LaNiO$_3$ layers in (111)-oriented LaNiO$_3$/LaMnO$_3$ superlattices probed by soft x-ray ARPES.

F. Y. Bruno[1], M. Gibert[1], S. McKeown Walker[1], O.E. Peil[1], A. de la Torre[1], S. Riccò[1], Z. Wang[1,2], S. Catalano[1], A. Tamai[1], F. Bisti[2], V. N. Strocov[2], J-M Triscone[1] and F. Baumberger[1,2]

[1] *Department of Quantum Matter Physics, University of Geneva, 24 Quai Ernest-Ansermet, 1211 Geneva 4, Switzerland*

[2] *Swiss Light Source, Paul Scherrer Institute, CH-5232 Villigen, Switzerland*

Taking advantage of the large electron escape depth of soft x-ray angle resolved photoemission spectroscopy we report electronic structure measurements of (111)-oriented [LaNiO$_3$/LaMnO$_3$] superlattices and LaNiO$_3$ epitaxial films. For thin films we observe a 3D Fermi surface with an electron pocket at the Brillouin zone center and hole pockets at the zone vertices. Superlattices with thick nickelate layers present a similar electronic structure. However, as the thickness of the LaNiO$_3$ is reduced the superlattices become insulating. These heterostructures do not show a marked redistribution of spectral weight in momentum space but exhibit a pseudogap of ≈ 50 meV.

## Main Text

With the advancement of synthesis techniques it is possible nowadays to control the growth of transition metal oxide (TMO) epitaxial heterostructures at the unit cell level. The realization of such heterostructures have resulted in the discovery of fascinating phenomena as a consequence of the confinement of strongly correlated electrons in ultrathin high quality layers. [1–3] Since these thin layers are embedded in heterostructures, interfacial phenomena such as charge transfer and magnetic coupling play an important role together with dimensionality in determining the ground state of the system. [4–8] The recent development of combined angle-resolved photoemission spectroscopy (ARPES) and oxide heterostructures growth setups has allowed the electronic structure of TMO heterostructures to be probed directly.

However, the small electron escape depth in VUV-ARPES experiments have limited most studies to the top layer of oxide thin films. [2,9–11]

Most members of the nickel based perovskites oxides $RNiO_3$, (R is a trivalent rare earth) exhibit a metal-to-insulator transition as a function of temperature. [12–15] In addition, a transition from a paramagnetic to an antiferromagnetic ground state is observed in these compounds. The critical temperature for both transitions is a function of the Ni-O-Ni bond angle and can be controlled by changing the R ion size. The least distorted member of the nickelates family, $LaNiO_3$ (LNO), constitutes an exception since it remains rhombohedral, metallic and paramagnetic at all temperatures. Despite presenting this rather simple bulk phase diagram, studies of thin LNO films have surged motivated by the possibility of tailoring their properties by epitaxial strain, [16,17] thickness control, [18–20] and particularly by using them as a building block in epitaxial oxide heterostructures and devices [3,21–29]. It has been shown by transport experiments that while thick (001)-LNO films grown on different substrates are metallic, as the thickness is reduced a metal to insulator transition is observed. [9,18] A similar dimensionality induced metal-insulator transition accompanied by the stabilization of an antiferromagnetic ground state was found in LAO/LNO superlattices. [3] Photoemission spectroscopy experiments have consistently reported a loss of spectral weight at the Fermi level ($E_F$) accompanying the thickness driven metal-insulator transition, however, there is no consensus of the underlying physical mechanisms behind this observation. [9,30–32] ARPES experiments showed that the 3D electronic structure of (001)-oriented LNO thin films remains intact down to 15 Å while for thinner films the destruction of Fermi liquid like quasiparticles [9] and the appearance of 1D Fermi surface (FS) nesting [32] was reported.

Interfacial and confinement effects in heterostructures grown along the perovskite - (111) orientation are much less studied [24,29,33,34]. Interestingly, in a thin film with a thickness of two monolayers (ML) of LNO grown along this orientation the Ni atoms form a buckled honeycomb lattice, such heterostructure was predicted to host novel topological phases which are yet to be found experimentally. [35–38] The recent realization of high quality heterostructures that combine LNO with thin insulating ferromagnetic $LaMnO_3$ (LMO) layers along the (111) direction and exhibit unexpected phenomena is an example of the novel physics found in this type of heterostructures.

Transport experiments reveal a metal to insulator transition in (111)-LNO/LMO superlattices as the LNO thickness is reduced. The magnetic properties of the insulating phase are rather surprising. When the samples are cooled down with an applied magnetic field they display negative exchange bias (EB). [24] The sign of the EB can be reversed as a function of temperature for heterostructures with an LNO thickness of 7 monolayers (MLs), defined as the distance between consecutive Ni atoms of ~ 2.19 Å. This behavior was interpreted based on the emergence of an AF spiral with a (1/4, 1/4, 1/4) wave vector in the nickelate layer that couples the ferromagnetic LMO layers. [28]

Here we employ soft x-ray ARPES (SX-ARPES) to study the electronic structure of metallic (111)-LNO thin films for the first time. The choice of this experimental technique was dictated by the larger photoelectron mean free path and thus bulk sensitivity compared to the conventional VUV-ARPES as well as concomitant sharp definition of the surface-perpendicular momentum $k_z$ [39,40]. The measured Fermi surface (FS) broadly agrees with previous experiments in (001) films and consists on an electron pocket at the Brillouin zone (BZ) center surrounded by eight hole pockets centered at the BZ corners. [9,10,41] We find that a simple two-band model of the Ni $e_g$ states assuming a cubic BZ does not fully describe the electronic structure whereas density functional theory (DFT) calculations including the bulk octahedral tilt pattern closely resemble the experimental Fermi surfaces providing a solid base for our observations. The large photon energies used in our study allowed us to map the electronic structure of LNO layers embedded in LNO/LMO superlattices overcoming the inherent surface sensitivity of the technique and demonstrating its effectiveness in studying interfacial phenomena. Finally we show that concomitant with the thickness induced transition to insulating behavior in transport experiments, a loss of spectral weight at $E_F$ is observed in heterostructures with 7 MLs thick LNO layers. However, no obvious sign of dimensional crossover or signatures of AF ordering are observed.

We have employed SX-ARPES to investigate the electronic structure of a LNO thin film and superlattices grown on STO-(111) oriented substrates. The experiments have been carried out at the SX-ARPES end station of the ADRESS beamline of the Swiss Light Source, (Switzerland). [39]. We used *p*-polarization of incident X-rays unless stated otherwise and the sample temperature was set to $T$=12 K. All samples were grown *ex-situ* by off-axis sputter deposition as described in Ref. [24] and exposed to air prior to the ARPES experiments. We first examined $LNO_{15}$, a thin film with a thickness of 15 MLs. Then, we proceeded to study the evolution of the electronic structure on two different $[LNO_N/LMO_M]_X$ heterostructures composed of X repetitions of a $LNO_N/LMO_M$ bilayer where M and N are the number of monolayers of each material. In both cases the superlattices are terminated in a top LMO insulating layer. We explored $[LNO_{15}/LMO_5]_4$ where the LNO thickness in each layer matches the thin film, and then we probed $[LNO_7/LMO_5]_6$ a sample with reduced nickelate thickness expected to show EB and a spin spiral with ordering vector along the (1/4, 1/4, 1/4) direction. Resistivity measurements as a function of temperature are shown in Fig. 1a where two different behaviors are observed. A weakly metallic behavior in the sense that $dR/dT>0$ is observed down to ≈150K and 50K for $LNO_{15}$ and $[LNO_{15}/LMO_5]_4$ samples. On the other hand the sample with a thinner nickelate layer $[LNO_7/LMO_5]_6$ is clearly insulating ($dR/dT<0$) consistent with the reported thickness induced metal-to-insulator transition. In Fig. 1b we show a magnetic hysteresis loop obtained at a temperature $T$=5K for the sample $[LNO_7/LMO_5]_6$ under zero field cooled (blue) and 0.05 T-field cooled (red) conditions. The 190 Oe shift in the field cooled hysteresis loop confirms the presence of EB in the sample, while this effect is not observed for $[LNO_{15}/LMO_5]_4$ with a thicker LNO layer (not shown). [24,28]

The results of our electronic structure measurements by means of SX-ARPES performed on $LNO_{15}$ are shown in Fig. 2. Fermi Surface (FS) measurements obtained with 500 and 382 eV excitation energy are displayed in Figs. 2a and 2b. These energies correspond to the high symmetry points Γ and R in normal emission for the cubic unit cell considered here. [39,42] In Fig. 2j we present a tight binding calculation of the bulk FS which consists of an electron pocket centered at Γ and hole pockets centered at the R points. The calculations assume a ratio between nearest-neighbor hopping ($t$) and next-nearest-neighbor hopping ($t'$) $t'/t$=0.13 known to be appropriate for describing the (001)-LNO FS [43]. In Figs. 2d and 2e we show the corresponding cuts of the tight-binding calculation through the high symmetry points in order to better compare with the experimental data. Electron pockets are shown in red and hole pockets in green. The experimental FS centered at Γ is formed by a central quasi hexagonal electron pocket surrounded by triangular hole pockets. These features are in fair agreement with the calculations. Dispersion plots measured along $k_{[1\,1\,-2]}$ direction obtained at different Γ points are shown in Figs. 2k and 2l. We observe that the band derived from the $e_g$ states of Ni which forms the electron pocket extends up to ~ 170 meV. A similar bandwidth was observed for thin films grown along the (001) direction indicating that distortions imposed by the substrate do not play a major part in determining this parameter. [44] We then turn our attention to the k-space cut through the R point at the BZ corner. In this measurement shown in Fig. 2b the electron pocket is absent as expected while large Fermi surface contours corresponding to cuts through the center of the bulk hole pockets are found at the locations predicted by the model in Fig. 2e. However the correspondence between calculation and data is not entirely satisfactory. The small hole contours predicted by the model are not resolved clearly in the experiment and the shape of the large hole contours differs slightly from the model. These subtle discrepancies could arise from the use of a simplified cubic unit cell in our tight binding model and/or from more intricate structural or electronic effects at the LNO/LMO interfaces. In order to distinguish these scenarios, we compare in Fig. 2g-2h our tight-binding model to a bulk DFT calculation that includes a trigonal distortion due to strain in the film and the bulk rotations of the $NiO_6$ octahedron. We first notice that the DFT-FS obtained for the Γ and R point are identical. This is due to the presence of backfolded bands, which are generally weak in ARPES. [45] In order to highlight the fundamental bands, we overlay the tight-binding model with a single atom in the basis. Interestingly we see that the hexagonal-like central hole pocket in Fig 2h is rotated 30 degrees in the DFT calculation with respect to the tight binding model. This is an effect of the octahedral rotations and in fact is observed in the data where we see that the sides of the hexagonal-like pockets are facing each other showing that the bulk-like rotations of LNO are not fully suppressed by the STO substrate.

In order to complete our characterization of the electronic structure of $LNO_{15}$ we scanned the excitation energy in the 350 - 650 eV range to study the $k_z$ dependence of the electronic structure. The results shown in Fig. 2c reveal a strong dispersion along $k_z$, characteristic of highly 3D electronic states. As is evident from a comparison with the model shown in Fig. 2f and 2i only half of the electronic bands are detected when measuring in this configuration. We confirmed that the bands are neither observed when measuring with *s*-polarized light suggesting an intricate matrix element effect. Taken together these measurements confirm the 3D nature of the electronic structure of the $LNO_{15}$ schematized by the tight binding model depicted in Fig. 2h, with the caveat that the shape of the hole pockets can only be correctly described if the bulk-like rotations of the NiO$_6$ octahedron are taken into account as in the DFT calculations shown in Fig 2h.

In Figs. 3a and 3b we show constant energy maps measured on the LMO terminated [LNO$_{15}$/LMO$_5$]$_4$ and [LNO$_7$/LMO$_5$]$_6$ heterostructures respectively obtained with 500 eV excitation energy. The electron inelastic mean free path for this energy is ~ 15Å which limits the probing depth of our study to the whole top LNO$_N$/LMO$_M$ bilayer. [46] The observed electronic structure is markedly similar for the [LNO$_{15}$/LMO$_5$]$_4$ sample and the LNO$_{15}$ film (cf. Fig. 2a). Even though the LNO thickness is the same in both layers, one would expect a modified FS as a consequence of the magnetic coupling and charge transfer at the interfaces [47]. However, the only appreciable effect in our data is a slight widening of the features in the spectra of the heterostructure, possibly due to the finite roughness at the interface with the LMO top layer. When the thickness of the LNO layer is further reduced down to 7 MLs there is a considerable reduction of spectral weight at the Fermi level consistent with the insulating character of the sample and the band-like features broaden significantly. Intriguingly though, the k-space map obtained by integrating spectra over a narrow window around E - E$_F$ = 50 meV does not show a marked redistribution of spectral weight as it is characteristic for nesting driven itinerant spin-density wave phases. [48,49] While these findings do not exclude AF ordering in LNO as it was invoked in Ref. 28 to explain the exchange bias observed in superlattices with 7 MLs LNO, [24,28] they suggest that the insulating behavior of thin LNO layers in LNO/LMO superlattices is triggered by a gradual loss of quasiparticle coherence, rather than a well-defined phase transition to a magnetically ordered state.

We further investigated the loss of spectral weight at the Fermi level by analyzing the angle integrated spectra of the three mentioned heterostructures. In the valence band spectra of the superlattices shown in Fig. 4a there is a distinct peak ~2.3 eV below the Fermi level that is absent in the thin films and corresponds to the top LMO layer. In the near- E$_F$ region we identify two peaks located ~0.1 and 1 eV below E$_F$. The optical gap of LMO has been reported between 0.6 up to 1 eV and previous XPS measurements revealed the onset of the density of states at 0.5 eV below the Fermi level. [50–53] Hence, we conclude that the peak located 0.1 eV below E$_F$ corresponds to $e_g$ states in Ni with the feature at 1 eV likely having contributions from both LMO and LNO. Figure 4b depicts the region of the spectra close to the Fermi level where the loss of spectral weight at $E_F$ for the [LNO$_7$/LMO$_5$]$_6$ sample is evident. In order to quantify this effect we show in Fig. 4c empirical fits using a spectral function consisting of a Lorentz shaped peak multiplied by a Fermi function and a Gaussian to take into account the instrument resolution. The best fit to the integrated spectra shown in the figure is obtained by considering an effective Fermi level at -50 meV. Similar fits performed for k-resolved spectra throughout the BZ do not show a significant variation of this energy scale. Again, this is untypical for a nesting driven instability of the Fermi surface where the gap size would be expected to correlate with the degree of nesting at the ordering wave vector. [54] Further considering the absence of a large redistribution of spectral weight in momentum space discussed previously, we thus interpret the suppressed low-energy spectral weight as a pseudogap in the sense that is not associated with a clear symmetry breaking. Such a pseudogap is fully consistent with the insulating character observed in transport and might arise from strong local correlations, possibly in conjunction with disorder due to interface roughness.

In summary we reported the ARPES Fermi surface of a LaNiO$_3$ thin film grown along the less-conventional (111) crystallographic direction. The observed 3D-FS consisting of an electron pocket at Γ and hole pockets centered at the R points of the BZ is qualitatively consistent with bulk models and previous observations in (001) films. When incorporated into [LNO/LMO] heterostructures the gross electronic structure of LNO changes only slightly with the spectral weight distribution conserving is 3D character. For the thinnest LNO layers our ARPES data show a loss of weight at the Fermi level indicative of a pseudogap opening and consistent with the observation of insulating behavior in transport experiments.

## Acknowledgments

This work was supported by the Swiss National Science Foundation through Division II (200021_146995), Ambizione grant (PZ00P2_161327) and NCCR MARVEL. The research leading to these results has received funding from the European Research Council under the European Union's Seventh Framework Program (FP7/2007-2013)/ERC Grant Agreement no. 319286 (Q-MAC).

## Figure Captions

**Figure 1.** Figure 1. (a) Resistivity as a function of temperature of [LNO$_{15}$/LMO$_5$]$_4$ and [LNO$_7$/LMO$_5$]$_6$ superlattices. (b) Magnetic hysteresis loop obtained in a [LNO$_7$/LMO$_5$]$_6$ heterostructure after cooling with 1T (red) and 0T (blue) magnetic field.

**Figure 2.** Electronic structure of a 15 MLs (111)-LaNiO$_3$ film. (a) Spectral weight at E$_F$ as a function of in-plane momentum. The data were taken at $h\nu$ = 500 eV corresponding to a spherical cut through *k*-space intersecting the $\Gamma_{004}$-point in normal emission. (b) Fermi surface map cutting through the bulk R$_{003}$-point, taken with $h\nu$ = 382 eV. (c) Fermi surface along $k_z$(d-f) and (g-i) Fermi surface plots based on tight binding and bulk DFT calculations, respectively. (j) Schematic of the bulk-Fermi surface of LaNiO$_3$ obtained from a tight binding model on a cubic Brillouin zone. The high symmetry points are indicated. (k) and (l) Dispersion plots obtained at the cuts through Γ indicated by dashed lines in (d). All the measurements were obtained at a temperature of T = 12 K.

**Figure 3.** Constant energy maps of (a) [LNO$_{15}$/LMO$_5$]$_4$ and (b) [LNO$_7$/LMO$_5$]$_6$ centered at Γ, the maps are obtained by integrating over ±15 meV at E$_F$ and E – E$_F$ = 50 meV respectively. All the measurements were obtained at a temperature of T = 12 K.

**Figure 4.** (a) Angle integrated spectra corresponding to LNO$_{15}$ (brown) [LNO$_{15}$/LMO$_5$]$_4$ (orange) and [LNO$_7$/LMO$_5$]$_6$ (blue). The photoemission intensities have been scaled to match each other in the 0.35-0.45 eV region where the signal is originated mostly from the LNO layer. (b) Near E$_F$ region and (c) data (blue dots) and fit (red) showing the presence of a 50 meV pseudogap. All the measurements were obtained at a temperature of T = 12 K with an excitation energy of 500 eV.

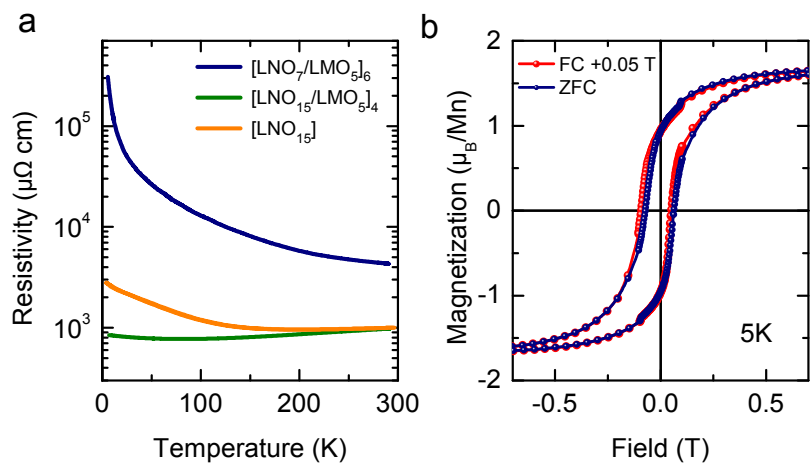

Figure 1. (a) Resistance as a function of temperature of [LNO15], [LNO$_{15}$/LMO$_5$]$_4$ and [LNO$_7$/LMO$_5$]$_6$ heterostructures. (b) Magnetic hysteresis loop obtained in a [LNO$_7$/LMO$_5$]$_6$ heterostructure after cooling with 1T (red) and 0T (blue) magnetic field.

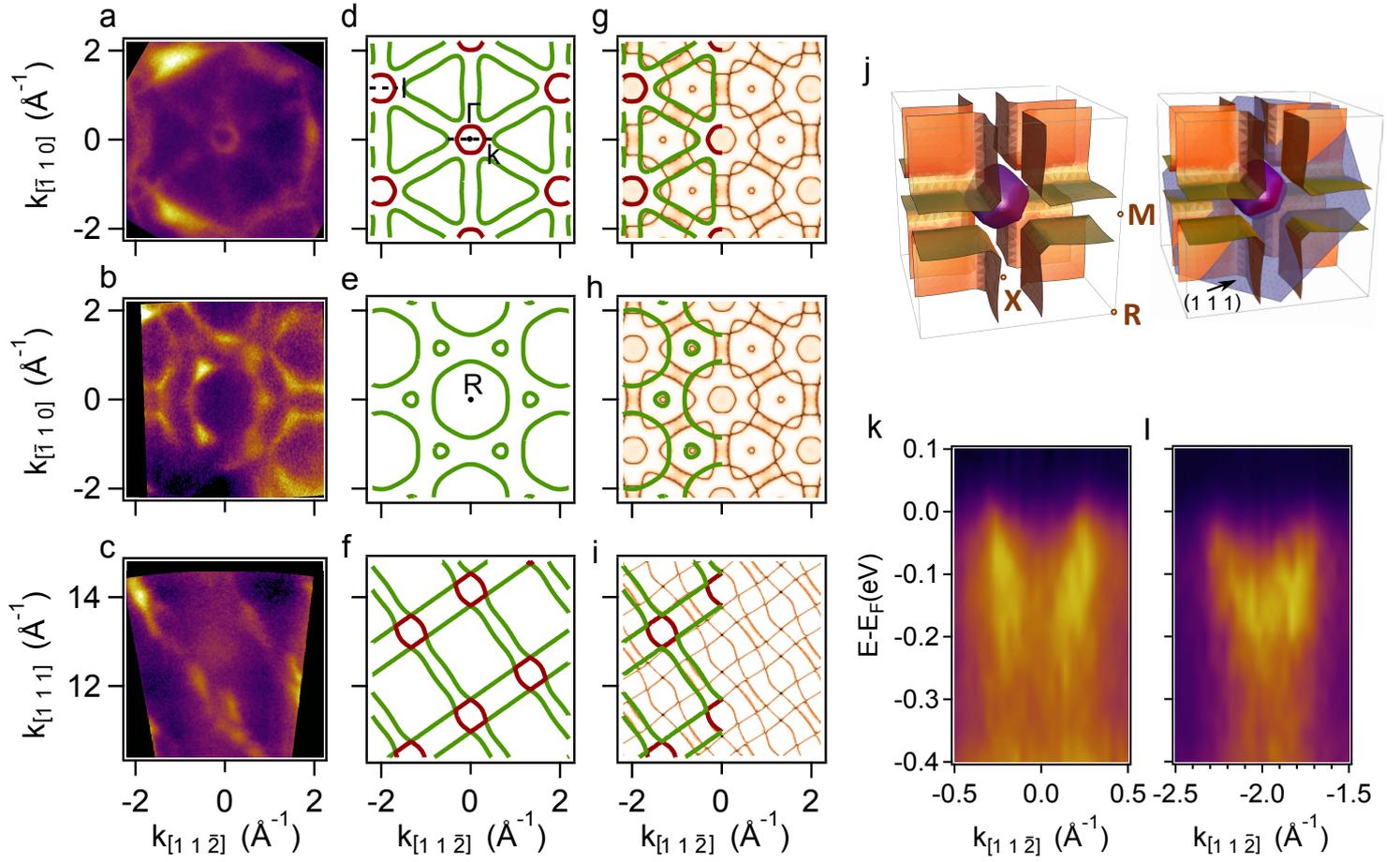

Figure 2. Electronic structure of a 15 MLs (111)-LaNiO$_3$ film. (a) Spectral weight at E$_F$ as a function of in-plane momentum. The data were taken at $h\nu$ = 500 eV corresponding to a spherical cut through k-space intersecting the $\Gamma_{004}$-point in normal emission. (b) Fermi surface map cutting through the bulk R$_{003}$-point, taken with h$\nu$ = 382 eV. (c) Fermi surface along k$_z$. (d-f) and (g-i) Fermi surface plots based on tight binding and DFT calculations. (j) Schematic of the bulk-Fermi surface of LaNiO$_3$ obtained from a tight binding model on a cubic Brillouin zone. The high symmetry points are indicated on the left and a (111) plane is shown on the right. (k) and (l) Dispersion plots obtained at the cuts through Γ indicated by dashed lines in (d). All the measurements were obtained at a temperature of T = 12 K.

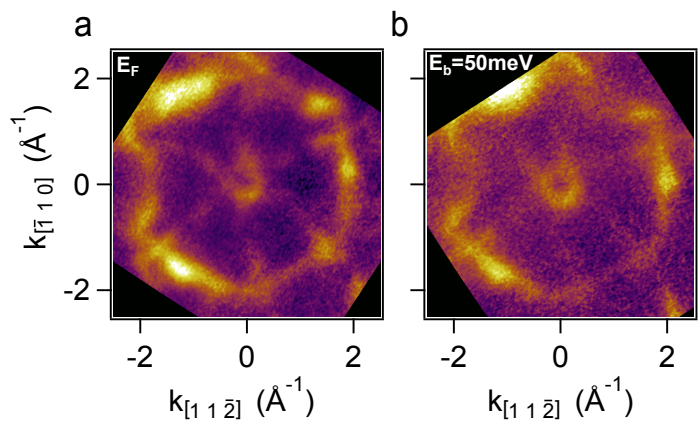

Figure 3. Constant energy maps of (a) $[LNO_{15}/LMO_5]_4$ and (b) $[LNO_7/LMO_5]_6$ centered at Γ, the maps are obtained by integrating over ±15 meV at $E_F$ and $E_b$ = 50 meV respectively. All the measurements were obtained at a temperature of T = 12 K.

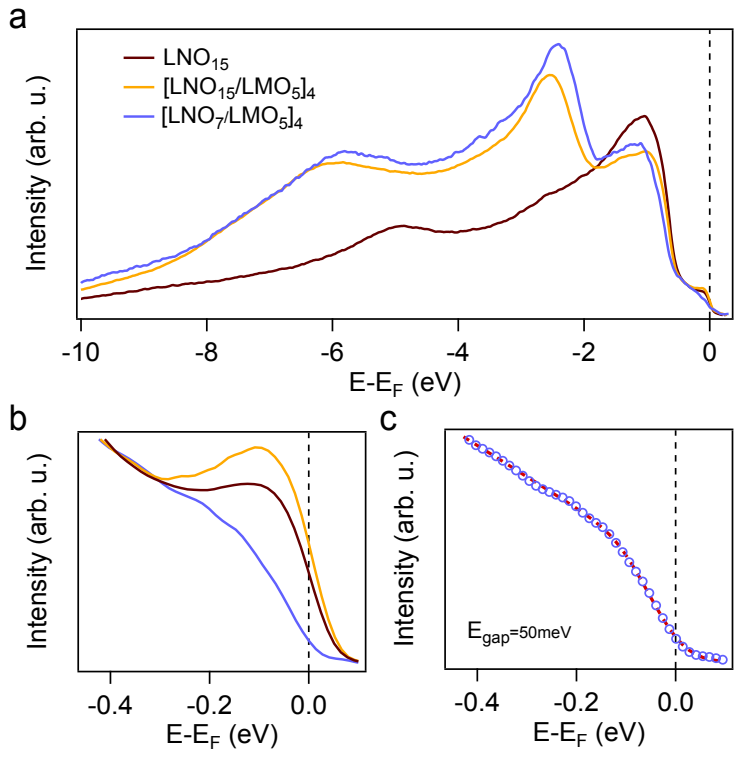

Figure 4. (a) Angle integrated spectra corresponding to $LNO_{15}$ (brown) $[LNO_{15}/LMO_5]_4$ (orange) and $[LNO_7/LMO_5]_6$ (blue). (b) Near $E_F$ region and (c) data (blue dots) and fit (red) showing the presence of a 50 meV pseudogap. All the measurements were obtained at a temperature T = 12 K with and excitation energy of 500 eV .

# Supplementary Information:

# Electronic structure of buried LaNiO₃ layers in (111)-oriented LaNiO₃/LaMnO₃ superlattices probed by soft x-ray ARPES.

F. Y. Bruno[1], M. Gibert[1], S. McKeown Walker[1], O.E. Peil[1], A. de la Torre[1], S. Riccò[1], Z. Wang[1,2], S. Catalano[1], A. Tamai[1], F. Bisti[2], V. N. Strocov[2], J-M Triscone[1] and F. Baumberger[1,2]

[1] Department of Quantum Matter Physics, University of Geneva, 24 Quai Ernest-Ansermet, 1211 Geneva 4, Switzerland

[2] Swiss Light Source, Paul Scherrer Institute, CH-5232 Villigen, Switzerland

**k_z determination**

For the photon energies used in this study, the free electron final state approximation is reliable and accurate. The momentum perpendicular to the surface $k_z$ can thus be determined from the relation $k_z = \sqrt{(2m_e/\hbar^2)(V_0 + E_k \cos^2\theta)} - k_\perp^{ph}$, where $E_k$ is the kinetic energy of the electron, $m_e$ the free electron mass and $\theta=0$ in normal emission. While the component of the photon momentum parallel to the sample plane was taken into account when converting angular to momentum scale in Figs 2a-2c, the perpendicular component $k_\perp^{ph}$ is much smaller and as a consequence was neglected. The inner potential V0 is determined self consistently from the ARPES data and the measured lattice parameters. The out of plane periodicity of our (111)-LaNiO₃ is c=2.19 Å as determined by x-ray diffraction. For this lattice parameter the Γ₀₀₄ and R₀₀₃ points are expected at $k_z$=11.48 Å⁻¹ and $k_z$=10.04 Å⁻¹, respectively. Inspection of various measurements as a function of photon energy lead as to conclude that Γ₀₀₄ and R₀₀₃ are probed with 500 eV and 382 eV, from which we derive an inner potential of V0 = 15 eV.

In addition to the $k_z$ dependence of the electronic structure of LNO₁₅ shown in the manuscript we show in Fig. S1 the measurements for [LNO₇/LMO₅]₆ obtained by varying the excitation energy in the 350 – 650 eV range. Even though there is a considerably widening of the features in the spectra, no obvious folding or additional periodicity is observed when compared to the electronic structure of the thin film (c.f. Fig. 2c).

**Absence of sample charging in [LNO$_7$/LMO$_5$]$_6$ spectra.**

Since the [LNO$_7$/LMO$_5$]$_6$ superlattice is insulating at low temperatures, we took special precautions to test for the possibility of sample charging. In Fig. S2 we show angle integrated spectra obtained for different sizes of the exit slit and thus different photon fluxes on the sample. As observed in the figure the spectra are basically coincident for different exit slit sizes which confirms the absence of sample charging. In the same figure we show a reference angle integrated spectrum measured on poly-crystalline Au at the same temperature and photon energy to illustrate the opening of a pseudogap.

**Density functional theory**

The DFT Fermi surfaces were obtained using Vienna Ab-initio Simulation Package (VASP) [1,2] within Generalized Gradient Approximation (PBE parametrization [3]). The plane-wave cut-off energy was set to 600 eV, a k-mesh of 16x16x16 points for the rhombohedral and of 20x20x20 points for the cubic structures was used to generate high-resolution eigenvalue sets in the Brillouin zone (BZ). Three sets of calculations were performed: an ideal cubic unit cell (space group Pm-3m), a bulk-like rhombohedral unit cell with R-3c space group, and a 111-supercell of the ideal cubic structure (i.e. an undistorted cubic structure defined as a rhombohedral unit cell with 2 Ni ions per unit cell) to identify folded Fermi sheets. Importantly, the R-point of the cubic unit cell $k_R = [ 1/2 , 1/2 , 1/2 ]$ turns into $k = [1, 1, 1]$ in the BZ of the rhombohedral unit cell, i.e. the R-cut is equivalent to the Γ-cut. As a result, these two cuts for the cubic unit cell are folded into one Γ-cut for the for the non-primitive rhombohedral unit cell as shown in Fig. S3.

To simulate the geometry of the strained thin film the rhombohedral unit cell of the bulk LNO was transformed to the hexagonal setting, with the a lattice constant chosen to match that of STO (a = 3.905 Å ·√2 ≈ 5.523 Å) and c lattice constant set to the value obtained from sample characterization by X-ray diffraction (c = 6 · 2.19 Å = 13.14 Å). When one transforms back to the trigonal setting the resulting R-3c unit cell turns out to be trigonally deformed along 111 direction compared to the bulk one, with the axis angle γ = 61.3° (compared to γ = 60.7° for the original bulk structure). This deformation is mainly responsible for the reduction of the small pockets visible in the R-cut of the cubic unit cell.

**Spherical vs. plane cuts**

In Figure S4 we show the difference between a FS obtained as cuts of the 3D electronic structure of LNO with a plane passing through the Γ point (a,c) and a spherical k-space contour probed by our SX-ARPES experiment (b,d). In the manuscript we show plane cuts since this is easier to interpret and describing the curved cuts will confuse the reader. We notice however that some of the distortions observed in the measurements can be explained based on these maps.

**References**

[1] G. Kresse and D. Joubert, Phys. Rev. B 59, 1758 (1999).

[2] G. Kresse and J. Furthmuller, Phys. Rev. B 54, 11169 (1996).

[3] J. P. Perdew, K. Burke, and M. Ernzerhof, Phys. Rev. Lett. 77, 3865 (1996).

**Supplementary Figure Captions.**

Figure S1. (a) and (b) ARPES Fermi surface of a $LNO_{15}$ and $[LNO_7/LMO_5]_4$ respectively. Note that the <111> direction is perpendicular to the sample surface.

Figure S2. Angle integrated spectra of $[LNO_7/LMO_5]_6$ obtained with different exit slit sizes (blue) and Au reference spectra (green).

Figure S3. Fermi surface of $LaNiO_3$ based on DFT band structure. (a) and (b) FS centered at Γ and R point respectively using an undistorted perovskite structure and a cubic unit cell. (c) FS centered Γ calculated using an undistorted perovskite structure given by a non-primitive rhombohedral unit cell.

Figure S4. (a,c) and (b,d) Fermi surface plots centered at the Γ point based on tight binding and DFT calculations respectively. (a,b) The plots are produced by cutting the 3D FS with a plane passing through the $Γ_{004}$ point. (c,d) the plots are produced by cutting the 3D FS with a spherical surface of radii k=11.49 A-1 corresponding to a photon energy of 500 eV.

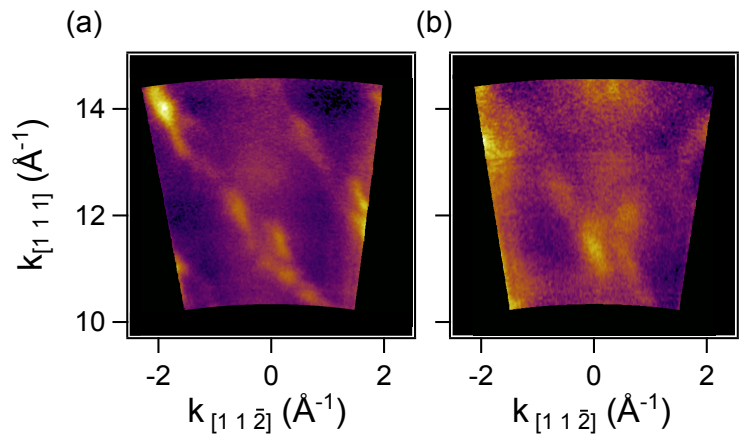

Figure S1. (a) and (b) ARPES Fermi surface of a LNO$_{15}$ and [LNO$_7$/LMO$_5$]$_4$ respectively. Note that the <111> direction is perpendicular to the sample surface.

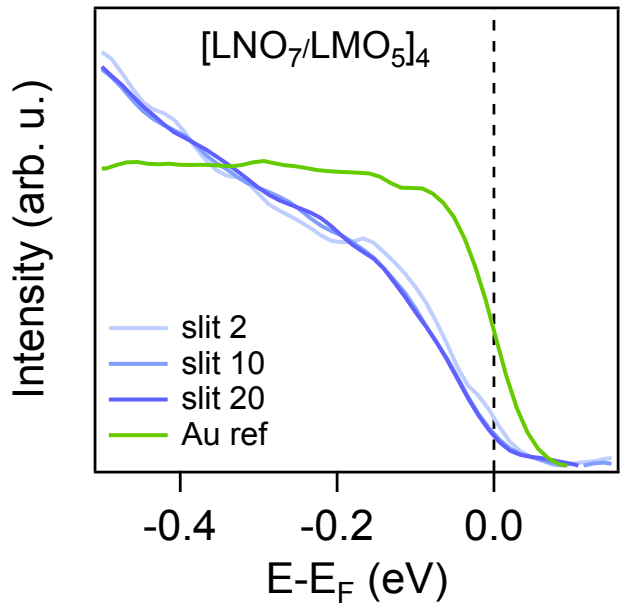

Figure S2. Angle integrated spectra of [LNO$_7$/LMO$_5$]$_6$ obtained with different exit slit sizes (blue) and Au reference spectra (green).

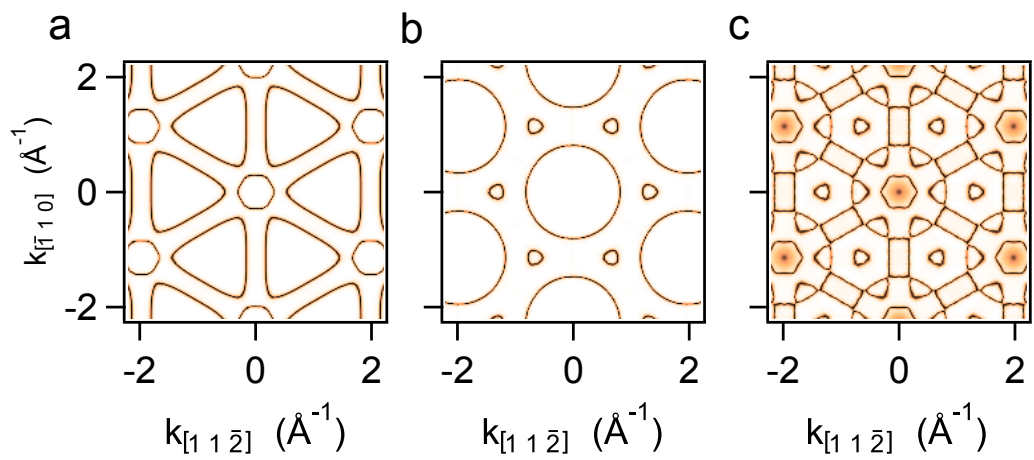

Figure S3. Fermi surface of LaNiO$_3$ based on DFT band structure. (a) and (b) FS centered at Γ and R point respectively using an undistorted perovskite structure and a cubic unit cell. (c) FS centered Γ calculated using an undistorted perovskite structure and a rhombohedral unit cell.

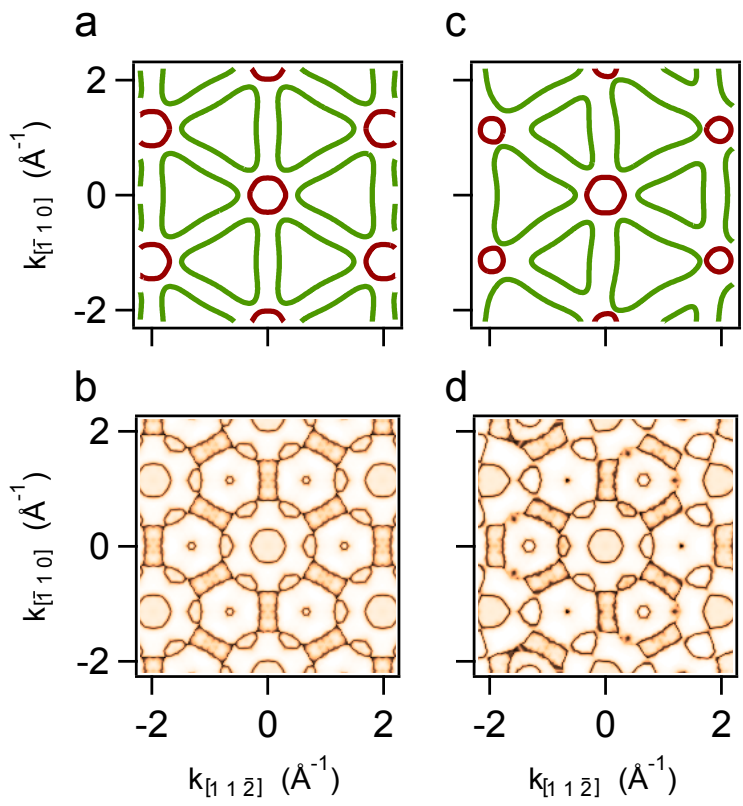

Figure S4. (a,c) and (b,d) Fermi surface plots centered at the Γ point based on tight binding and DFT calculations respectively. (a,b) The plots are produced by cutting the 3D FS with a plane passing through the $\Gamma_{004}$ point. (c,d) the plots are produced by cutting the 3D FS with a spherical surface of radii k=11.49 Å$^{-1}$ corresponding to a photon energy of 500 eV.